\documentclass[11pt]{article}
\usepackage{fa2025}
\usepackage{amsmath}
\usepackage{url}
\usepackage{graphicx}
\usepackage{color}
\usepackage{siunitx}
\usepackage[utf8]{inputenc}
\usepackage[backend=biber,uniquelist=false,maxbibnames=9,maxcitenames=2,mincitenames=1,style=ieee, 
url=false, isbn=false, eprint=false,
doi=false,
natbib=true]{biblatex}
\usepackage{multirow}
\usepackage{caption}
\usepackage{subcaption}

\AtEveryBibitem{
  \clearlist{language}
}
\AtEveryBibitem{\clearfield{month}}
\AtEveryBibitem{\clearfield{day}}
\AtEveryBibitem{\clearfield{publisher}}

\usepackage{csquotes}
\usepackage[inline]{enumitem}
\definecolor{red}{RGB}{0,0,0} 

\title{Tracking of Intermittent and Moving Speakers: Dataset and Metrics}

\multauthor
{Taous Iatariene$^{12*}$, Alexandre Guérin$^1$, Romain Serizel$^2$} {\\
  $^1$ Orange Innovation, 35000 Rennes, France\\
 $^2$ Université de Lorraine, CNRS, Inria, LORIA, 54000 Nancy, France 
 \correspondingauthor{taous.iatariene@orange.com}{Taous Iatariene et al.}
}

\begin{document}

\maketitle

\begin{abstract}
This paper presents the problem of tracking intermittent and moving sources, i.e, sources that may change position when they are inactive.
This issue is seldom explored, and most current tracking methods rely on spatial observations for track identity management. 
They are either based on a previous localization step, or designed to perform joint localization and tracking by predicting ordered position estimates.
This raises concerns about whether such methods can maintain reliable track identity assignment performance when dealing with discontinuous spatial tracks, which may be caused by a change of direction during silence.
We introduce LibriJump, a novel dataset of acoustic scenes in the First Order Ambisonics format focusing on speaker tracking. The dataset contains speakers with changing positions during inactivity periods, thus simulating discontinuous tracks.
To measure the identity assignment performance, we propose to use tracking association metrics adapted from the computer vision community.
We provide experiments showing the complementarity of association metrics with previously used tracking metrics, given continuous and discontinuous spatial tracks.
\end{abstract}

\section{Introduction}\label{sec:introduction}

Sound source tracking is a subtask of acoustic scene {analysis}, which aims at estimating the spatial tracks of sound sources from multichannel audio recordings. 
It involves two main considerations: spatial, mainly through Direction of Arrival (DoA) estimation,
and identity-related through track identity (ID) management techniques.
The last decade has demonstrated interest in this research field, as shown by the organization of the \hbox{LOCATA} challenge~\cite{eversLOCATAChallengeAcoustic2020}, where
efforts were made towards the construction of an open recorded dataset of moving speakers in various configurations~\cite{lollmannLOCATAChallengeData2018}, and the sharing of an evaluation framework to evaluate tracking systems\footnote{\url{https://github.com/cevers/sap_locata_eval}}.

{Most studies focus on tracking continuously active sources, setting aside the challenging yet realistic scenario of intermittent and moving sound sources, which may move while silent.
To the best of our knowledge, only seldom works
have addressed this issue, often by tracking both active and inactive sources and assuming continuous movement during inactivity~\cite{quinlanTrackingIntermittentlySpeaking2009,liOnlineLocalizationTracking2019}.
However, difficulties arise when change of direction occurs during silence and lead to discontinuous spatial tracks~\cite{liOnlineLocalizationTracking2019}. Indeed, track ID management in such cases becomes more challenging, increasing the risk of assigning multiple redundant identities to the same source.}
Another resulting issue lies in the careful choice of evaluation metrics for evaluating track ID assignment, which is the principal type of error induced by movement during silences.

In this paper, we focus on speaker tracking and propose the following contributions. 
First, we introduce \hbox{LibriJump}, an evaluation dataset of synthetic acoustic scenes in the First Order Ambisonics (FOA) format containing intermittent and moving speakers. In this dataset, speakers remain static while speaking but change position (jump) when silent, leading to discontinuous spatial tracks. 
Second, we propose to complement the \hbox{LOCATA} tracking metrics with the tracking association metrics~\cite{luitenHOTAHigherOrder2021}, which are focused on evaluating the ID assignment performances of tracking systems on a global level rather than on a short context (frame) level. Initially proposed \hbox{by~\citet{luitenHOTAHigherOrder2021}}, these metrics have gained popularity in Multi-Object-Tracking.
\textcolor{red}{Finally, we analyze the complementarity of the proposed global association metrics with frame-level identity related tracking metrics.}


\section{Problem formulation}


We consider an indoor acoustic scene in which one or several speakers are discussing, such as an office meeting between colleagues. In such scenarios, speakers may move around, either while speaking or when silent. 
Given a {microphone} array (multichannel) recording of such acoustic scene, the speaker's positions relatively to the microphone array (i.e, the DoA), can be extracted. 

\begin{figure}[!b]
    \centering
    \includegraphics[width=0.9\linewidth]{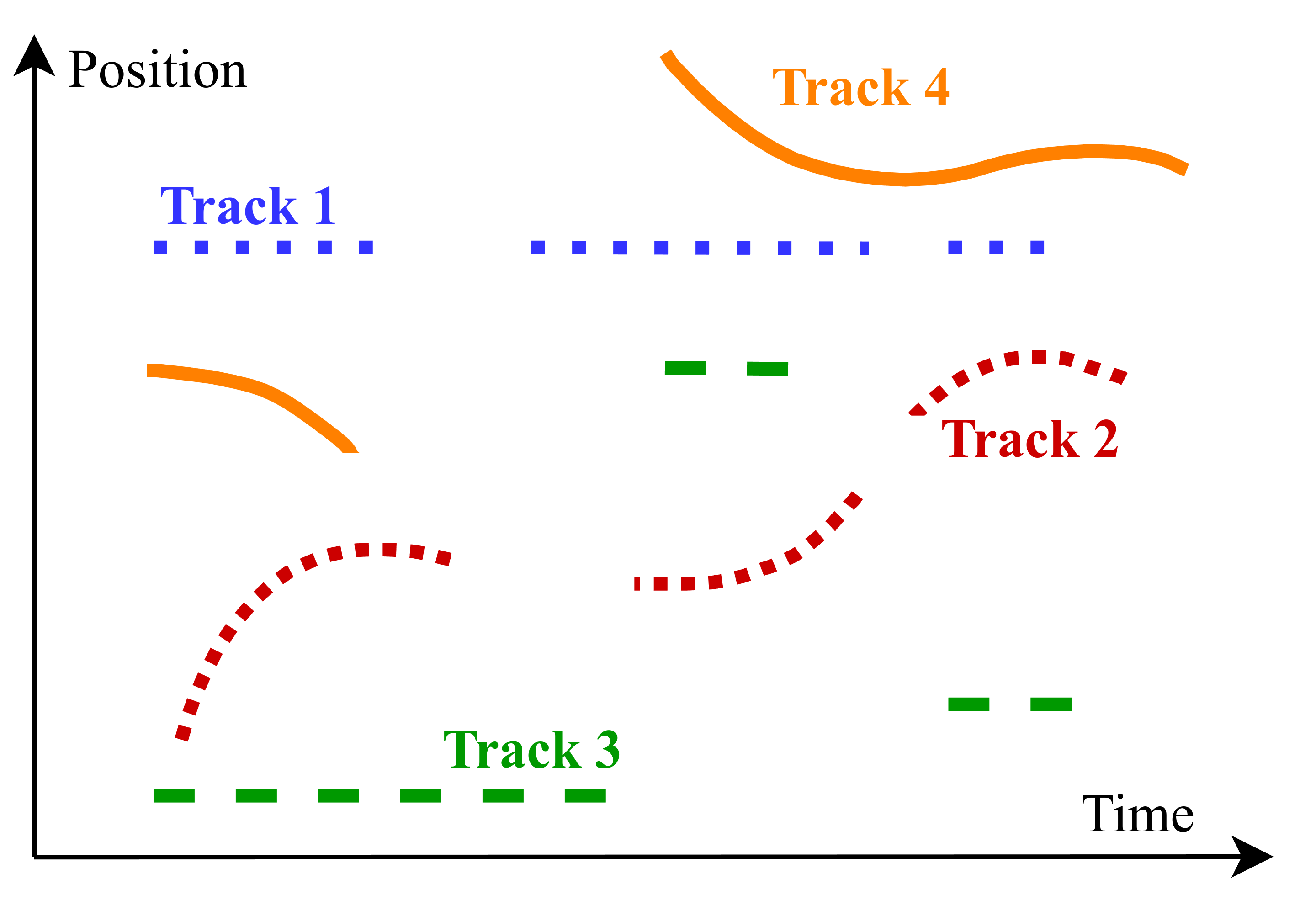}
    \caption{{Example of continuous and discontinuous tracks. Each color designs a different track ID.}}
    \label{fig:pb_formulation}
\end{figure}

In the rest of this paper, we denote as $J$ the number of speakers in the scene. We use the term \enquote{track} to refer to the time-varying DoA of each speaker.
As illustrated by Fig.~\ref{fig:pb_formulation}, each track is labeled with a track identity (ID) designing its corresponding speaker.
Depending on the nature of movement during silence, a track can be either continuous or discontinuous. 
Continuous tracks occur when the speaker is static (Track 1 in Fig.~\ref{fig:pb_formulation}) or when it moves continuously when silent (Track 2).
Discontinuous tracks can be due to non-predictable movement occurring during silence, whether the speaker is static while speaking (Track 3), or moving (Track 4).

The goal of a tracking system is to reconstruct the speakers' spatial tracks given the multichannel recording. This implies two main considerations: spatial DoA estimation, and track ID management (i.e, assigning the correct ID to DoAsthrough time).
In the rest of this paper, we use the term \enquote{predictions} or \enquote{predicted tracks} to refer to the output of a tracking system.
Similarly, \enquote{ground truths} or \enquote{ground truth tracks} will refer to the real speaker's time-varying position. 

\section{Related works}

\subsection{Sound source tracking} 
\label{sec:review_methods}


A two-step approach of sound source localization for DoA estimation~\cite{grumiauxSurveySoundSource2022}, followed by Bayesian filtering to link the DoAs through time, is broadly adopted for tracking
\cite{eversLOCATAChallengeAcoustic2020}.
In such approaches, tracks' \enquote{states} (position, velocity) are updated at each new timestep given a new set of DoA observations, using a Bayesian filter like a particle filter~\cite{kiticTRAMPTrackingRealtime2018a}.

In the more challenging multi-speaker case, observation-to-track association is usually performed before Bayesian filtering.
Track ID management rely generally on the latter step, in which tracks are initiated, updated or deleted given the {result} of observation-to-track association. 
{Thus, we observe that track ID management relies on DoAs (spatial observations).} While this may be acceptable when {speakers} are continuously active or when their movement while silent is predictable (tracks 1 and 2 on Fig.~\ref{fig:pb_formulation}), ID assignment errors may occur, especially in case of 
\textcolor{red}{change of direction} during silence like tracks 3 and 4 in Fig.~\ref{fig:pb_formulation}. 

Only several works have addressed tracking of intermittent and moving speakers.{~\citet{quinlanTrackingIntermittentlySpeaking2009, liOnlineLocalizationTracking2019}} proposed to track a fixed number of simultaneously active speakers, whether active or inactive, by assuming continuous movement during silence for track position update.~\citet{quinlanTrackingIntermittentlySpeaking2009} proposed to estimate dynamically the number of active speakers $N_a$ and mark as active the $N_a$ tracks with the highest probability of existence.~\citet{liOnlineLocalizationTracking2019} proposed to evaluate the speech activity of each track. 
\textcolor{red}{Discontinuous tracks due to direction changes during silences were a limitation of their study.}

Recent deep learning-based tracking approaches were proposed, by predicting ordered position estimates, through the permutation-aware training of localization networks~\cite{adavanneDifferentiableTrackingBasedTraining2021,diaz-guerraPositionTrackingVarying2023a}.
Emphasis has been given on the spatial track reconstruction, setting aside the track ID management aspect. The number of predicted tracks is limited by the number of output branches of the networks. Also, the permutation-aware training usually lead to the \textit{block permutation problem} described 
in speech separation~\cite{chenContinuousSpeechSeparation2021}, 
but still not addressed for deep-learning based audio tracking.

We recently proposed a novel approach, focused on speaker tracking, where DoAs and identity-related observations (speaker embeddings) were combined to improve the identity assignment performance of several tracking systems, in the occurring of discontinuous tracks~\cite{iatarieneSpeakerEmbeddingsImprove2025}.


\subsection{Moving speakers datasets}

We focus on this section on summarizing the existing datasets of moving speakers. 
A more exhaustive overview of existing sound source tracking datasets can be found in the works of~\citet{grumiauxSurveySoundSource2022, yangRealMANRealRecordedAnnotated2024}.

Due to the difficulty of acquiring metadata of the sources' positions, which is crucial for supervised training and evaluation of tracking systems, only a few datasets of real recordings are available. 
This is why it is often resorted to synthetic data generation, where room acoustics (spatialization and reverberation) is simulated through Spatial Room Impulse Responses (SRIRs), which are then convolved with so-called \textit{dry} (clean) speech sources. We note in particular the works of~\citet{
diaz-guerraRobustSoundSource2021}
to simulate acoustic scenes with moving sound sources. 

As for labeled recorded datasets of moving speakers, the \hbox{LOCATA} data corpus contains recordings (\hbox{$\approx 1$ hour}) of static and moving english speakers, using multiple microphone arrays~\cite{lollmannLOCATAChallengeData2018}.
The recent \hbox{RealMAN} dataset, is a larger scale recorded corpus (83~hours) of static and moving mandarin speakers, in various acoustic scenes~\cite{yangRealMANRealRecordedAnnotated2024}.

All previously mentioned datasets provide recordings of continuously moving sources (whether static or moving). 
As a result, recordings containing speakers with discontinuous spatial tracks are scarce to find. 

\subsection{Tracking metrics}

Tracking systems require metrics evaluating the two aforementioned aspects of tracking: spatial-related through localization metrics, and identity-related through the measurement of track ID assignment errors. 
The \hbox{LOCATA} challenge have provided a first set of metrics derived from Multi-Target-Tracking~\cite{eversLOCATAChallengeAcoustic2020}.
\textcolor{red}{The CLEAR metrics from Multi-Object-Tracking (MOT)~\cite{bernardinEvaluatingMultipleObject2008} have also been adapted to audio tracking more recently
~\cite{adavanneDifferentiableTrackingBasedTraining2021}}.
We provide here insights on the computation of those metrics.

Given ground truths and predicted tracks, a prior to metric computation is to perform a one-to-one matching between predictions and ground truths, on the frame level. 
Matching pairs form \hbox{True Positives~(TPs)}, while non-assigned predictions become \hbox{False Positives~(FPs),} and non-assigned ground truths \hbox{False Negatives~(FNs)}.

Localization metrics are estimated on the set of TPs, often using the mean angular distance 
as an error measure~\cite{adavanneDifferentiableTrackingBasedTraining2021,diaz-guerraPositionTrackingVarying2023a}.
For identity related metrics, further distinction can be made between measuring detection errors (taking into account FPs and FNs) and association errors (ID mismatches within the set of TPs). We focus in the following on association metrics.

In \hbox{LOCATA}, track ID \textit{swaps} on the set of TPs are {defined}, by comparing the consistency of the predicted track IDs given their matching ground truth track IDs, from frame to frame. Any change in predicted ID given the same ground truth ID result in a swap. 
\textit{Broken} tracks 
are also measured
by counting the number of times a ground truth track is associated then remains unassociated on the next frame (i.e, moving from TP to FN). 
Track Swap Rate ({TSR}) is defined as the number of swapped tracks, and Track Fragmentation Rate ({TFR}) as the number of broken plus swapped tracks.
The CLEAR metrics defines Identity Switches ({IDSW}) similarly to track ID \textit{swaps}. 

Global metrics combining localization, detection and association errors, have been developed to evaluate the overall tracking efficiency.
In \hbox{LOCATA}, the Optimal SubPattern Assignment (OSPA) metric combines cardinality error 
and localization error,
making it able to evaluate both detection and localization.
The CLEAR metrics propose the Multi Object Tracking Accuracy (MOTA) score, in which IDSW is combined with FNs and FPs, thus measuring detection and association errors.

\section{\hbox{LibriJump} dataset}

We present \hbox{LibriJump}, an evaluation dataset containing intermittent and moving speakers\footnote{The dataset will be shared on \url{zenodo.org} upon publication.}. 

\subsection{Description}

LibriJump is a synthetic dataset, divided into three subsets (\hbox{LibriJump}-1spk, \hbox{LibriJump}-2spk, \hbox{LibriJump}-3spk) for three different values of $J$ (the number of speakers in the scenes). 
Each subset contains 150 mixtures of acoustic scenes in the First Order Ambisonics (FOA) format, of 60 second each.
The main objective of this dataset is to provide the scientific community with acoustic scenes containing speakers with {discontinuous spatial tracks}, in order to draw attention to the challenges raised by such scenarios in term of track ID assignment.
In \hbox{LibriJump}, emphasis is placed on track discontinuity, and challenges induced by spatially dynamic tracks (i.e, movement during activity) are minimized. Indeed, the latter scenario is already addressed in previously mentioned datasets.
Consequently, \textcolor{red}{similar to track 3 in Fig.~\ref{fig:pb_formulation}}, speakers remain static while speaking but move when silent, thus {simulating} \enquote{jumps} in speakers positions. 
\textcolor{red}{This allows easier 
spatial reconstruction, without taking anything away from the difficulty induced by discontinuities during silence in terms of identity assignment.}
\hbox{LibriJump} is derived from the LibriSpeech test-clean dataset~\cite{panayotovLibrispeechASRCorpus2015}.

\subsection{Methodology}
\label{sec:librijump_method}
To simulate a mixture of $J$ speakers, we: select \textit{dry} (clean) speeches, generate SRIRs, spatialize the dry speech, and add background noise. 

Dry speech sources are obtained by first selecting $J$ random speakers from the LibriSpeech test-clean dataset. For each speaker, several utterances are chosen randomly, and concatenated to reach 60 second of dry speech.

We obtain SRIRs through the simulation of \textit{Shoeboxes} rooms. For each shoebox room, 6 SRIRs are {generated}, corresponding to 6 positions relatively to one microphone position.
Positions are chosen randomly to achieve a minimal angular distance of 60° between each pair of {positions}.
We rely on a custom FOA SRIR {generator} based on Pyroomacoustics~\cite{scheiblerPyroomacousticsPythonPackage2018}.
Shoeboxes are generated by varying randomly the reverberation time (RT60) from 0.2 to 0.8 seconds, and the length and width from 2 to 10 m, and height from 2 to 3 m.

Each speaker is then spatialized individually. 
First, its corresponding dry speech is segmented using a voice activity detector\footnote{\url{https://huggingface.co/speechbrain/vad-crdnn-libriparty}}.
Each segment of speech is then convolved with one of the 6 SRIRs to simulated changing positions from one period of activity to another. We add a random silence from 0.1 to 1 second between each segment to reduce overlap.
Then, we adjust the level of one arbitrary speaker, and set the levels of the other speakers to be 2 to 6 dB lower.
The summation of the $J$ spatialized signals leads to the obtention of the unnoisy mixture.

A background noise is randomly selected from samples taken from Freesound~\cite{fontFreesoundTechnicalDemo2013}. It is then convolved with the diffuse part of two SRIRs to obtain a diffuse (non spatialized) noise.
The resulting diffuse noise is added to the mixture with a fixed SNR of 15 dB.

\section{Tracking association metrics}

\subsection{Description}
As with the CLEAR metrics, we propose to adapt to audio tracking the recent Higher Order Tracking Accuracy (HOTA) metrics proposed by~\citet{luitenHOTAHigherOrder2021}.
The HOTA metrics are divided into sub-metrics evaluating localization, detection and association errors.

We adapt the subset of HOTA measuring association errors: association accuracy, precision and recall (AssA, AssPr, AssRe). 
These association metrics provide a global evaluation of predictions against ground truths in term of track ID assignment, in contrast with IDSW, TFR and TSR which are measured at the frame level.

Tracking association metrics overcome numerous limits of CLEAR tracking metrics. As mentioned, IDSW evaluates association errors at the frame-level. This can penalize a tracker that has wrongly switched but corrects itself later on. 
Moreover, IDSW only focuses on the switches made on predicted IDs, setting asside the {evaluation} of switches that can be made over ground truth IDs.
The MOTA score values equally FP, FNs (related to detection) and IDSW (related to association). As a result, MOTA gives twice as much priority to detection evaluation than association.
More limits to CLEAR metrics can be found in the paper {of~\citet{luitenHOTAHigherOrder2021}}.

One limitation to association metrics is that it is not evaluating the degree of track fragmentation. 
Although a 
fragmentation aware HOTA
was proposed~\cite{luitenHOTAHigherOrder2021}, we choose to keep TFR to evaluate the degree of track fragmentation. 
Tracking association metrics are thus viewed as complementary to identity-related metrics.

\textcolor{red}{AssRe focuses on predicted IDs errors, and measures how a single ground truth track may be 
spread
into multiple predicted tracks, with a low score indicating a tendency to split a ground truth track into many predicted ones. 
AssPr focuses on ground truth IDs errors (not measured by IDSW), and measures how multiple ground truth tracks can be merged into a single predicted track, with a low score indicating that 
a single predicted track is an aggregate of multiple portions of ground truth tracks.
AssA provides an overall score combining AssRe and AssPr.}

\subsection{Calculation}

In addition to the previously defined concepts of TP, FP, FN,~\citet{luitenHOTAHigherOrder2021} define a novel concept of TPA, FPA and FNA ("A" standing for Association). 

More precisely, given a TP couple $c=(\text{prID},$ $\text{gtID})$ denoting a matched prediction and ground truth ID, respectively, 
TPAs are the TPs sharing the same ID match as~$c$ (i.e, the matched prID and gtID are the same). 
FPAs are the TPs couples with same prID, but with a different gtID, added to the FPs with same prID.
Similarly, FNAs are the TPs with same gtID, but different prID, added to FNs having same gtID.
\textcolor{red}{The reader is referred to Fig. 2 in~\cite{luitenHOTAHigherOrder2021} for a visual representation of TPA, FPA and FNA.}

AssA, AssPr, AssRe are then obtained by combining TPAs, FPAs and FNAs scores and averaging over TPs, as proposed by~\citet{luitenHOTAHigherOrder2021}:
\begin{equation}
    \label{eq:assre}
    \begin{aligned}
\operatorname{AssRe} & =\frac{1}{|\mathrm{TP}|} \sum_{c \in\{\mathrm{TP}\}} \frac{|\operatorname{TPA}(c)|}{|\operatorname{TPA}(c)|+|\mathrm{FNA}(c)|}
\end{aligned}
\end{equation}
\begin{equation}
    \label{eq:asspr}
    \begin{aligned}
\operatorname{AssPr} & =\frac{1}{|\mathrm{TP}|} \sum_{c \in\{\mathrm{TP}\}} \frac{|\operatorname{TPA}(c)|}{|\operatorname{TPA}(c)|+|\mathrm{FPA}(c)|}
\end{aligned}
\end{equation}
\begin{equation}
    \label{eq:assa}
    \begin{aligned}
\operatorname{AssA} & = \frac{1}{|\mathrm{TP}|} \sum_{c \in\{\mathrm{TP}\}} \mathcal{A}(c)
\end{aligned}
\end{equation}
\begin{equation}
\begin{aligned}
\text{with } \ \mathcal{A}(c) & =  \frac{|\operatorname{TPA}(c)|}{|\operatorname{TPA}(c)|+|\mathrm{FNA}(c)|+|\mathrm{FPA}(c)|} 
\end{aligned}
\end{equation}
where $|.|$ stands for the cardinal of the considered set.

\subsection{Adaptation to audio tracking}

The previous definitions of AssA, AssPr and AssRe are universal to any tracking system, as long as matching between ground truth and predicted tracks has been performed to obtain as set of TP, FP and FN. 
As such, the main difference between the initial HOTA metrics and our adaptation lies in the similarity measure used to compute the matches between predictions and ground truths: the Intersection-over-Union (IoU) used in MOT have been replaced by the angular distance for sound source tracking.



\section{Experiments}

We propose experiments to highlight the complementarity of the proposed tracking association metrics with previous identity related metrics. We use two tracking systems on two datasets (\hbox{LOCATA} and \hbox{LibriJump}).




\subsection{Tracking systems}

The tracking systems we use in the experiments reflect the diversity of the aforementioned tracking approaches: a Bayesian particle filter (PF)~\cite{kiticTRAMPTrackingRealtime2018a}, and a neural tracker (NN) adapted from the works of~\citet{subramanianDeepLearningBased2022}.

\subsubsection{Bayesian tracker}

The PF tracker consist in a two-step localization and particle filtering tracker.
Localization is performed using a neural network trained to estimate for each frame azimuth and elevation angles on a discrete spherical grid, given FOA input features~\cite{grumiauxImprovedFeatureExtraction2021}.
The particle filter's track ID management is based on a policy that takes into account the maximum number of simultaneously active sources (equal to the number of speakers in the scene $J$) and the maximum number of track IDs, noted $K_{max}$, which can be unbounded ($K_{max} = \inf)$~\cite{kiticTRAMPTrackingRealtime2018a}.
Two different hyperparameter sets are used, depending on the dataset (\hbox{LOCATA} or \hbox{LibriJump}) over which tracking is performed.


\subsubsection{Neural tracker}

The NN tracker is a neural network inspired from the Source Splitting approach of~\citet{subramanianDeepLearningBased2022}. It also predicts ordered DoAs on a discrete spherical grid, using the Permutation Invariant Training (PIT) strategy. 
It is designed with two output branches, which fixes the maximum number of predictable IDs to $K_{max} = 2$. 

\subsection{Datasets and metrics}

The dataset used for training the localizer and NN tracker described in the previous section is formed similarly to what is described in Sec.~\ref{sec:librijump_method}, except, notably, for the use of separate subsets for SRIRs, noise and speech (LibriSpeech train-clean-100 and train-clean-360), the use of a random SNR (between 5 and 50 dB).

In addition to \hbox{LibriJump}, we use real recordings from Tasks 1 and 3 of the \hbox{LOCATA} evaluation dataset, containing several single speaker scenes, either static (Task 1) or moving (Task 3). In those recordings, all spatial tracks are continuous, similar to tracks 1 and 2 of Fig.\ref{fig:pb_formulation}.
To simulate spatial discontinuity, we create a subset 3*, 
by zeroing some voice activity periods on Task 3 recordings.

We focus on identity related metrics, and use TFR and TSR~[$s^{-1}$], and the proposed AssPr, AssRe, AssA~[\%].
For LirbiJump dataset, mean scores with~80-20\% bootstrap are displayed, and standard deviation was below~1\%.
For \hbox{LOCATA} dataset, no bootstrap is performed due to the small number of recordings (13 for Task~1, 5 for Task~3).

\begin{figure*}[!htbp]
    \centering
    \begin{subfigure}{0.9\columnwidth}
        \centering
        \includegraphics[width=0.9\textwidth]{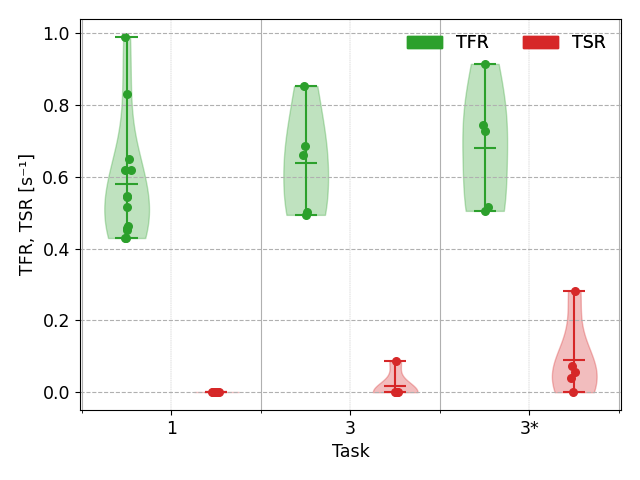}
        \caption{TFR and TSR $\downarrow$ [$s^{-1}$]}
        \label{fig:exp1_tfr_tsr}
    \end{subfigure}
    \begin{subfigure}{0.9\columnwidth}
        \centering
        \includegraphics[width=0.9\textwidth]{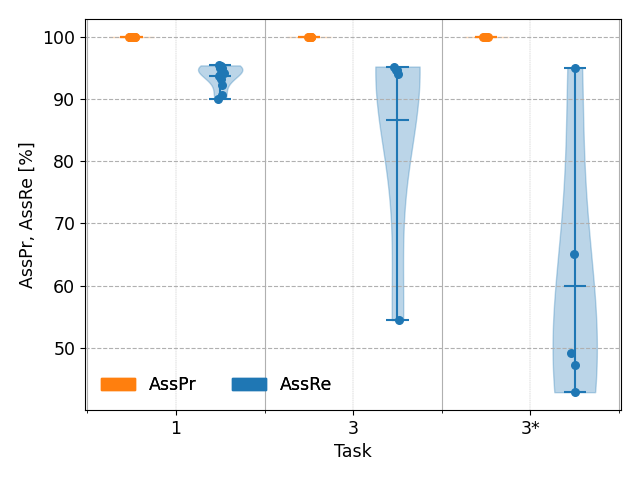}
        \caption{AssPr and AssRe $\uparrow$ [\%]}
        \label{fig:exp1_asspr_assre}
    \end{subfigure}
    \caption{Identity tracking metrics on Task 1 and 3 (single speaker) of \hbox{LOCATA} evaluation dataset, using PF tracker ($K_{max}=\inf)$. 
    {3*} refers to Task 3 with voice activity periods set to zero to induce discontinuity.}
    \label{fig:exp1}
\end{figure*}

\subsection{Results}

Metric scores using PF tracker on \hbox{LOCATA} (setting $K_{max}=\inf$) and \hbox{LibriJump} ($K_{max}=J,  \ 2J, \ 4J, \ \inf$) datasets are displayed on Fig.~\ref{fig:exp1} and Fig.~\ref{fig:exp2}, respectively.
Tab.~\ref{tab:tab_exp2} shows NN tracker performance on \hbox{LibriJump} ($K_{max} = 2$ fixed).
We first focus on single speaker scenes (\hbox{LOCATA} and \hbox{LibriJump}-1spk), then analyze identity performance in the more complex multi-speaker case (subsets 2spk and 3spk).


\subsubsection{Single speaker scenes analysis}

We {analyze} the results using TFR and TSR, then with association metrics to show their complementarity.

\begin{figure*}[!htbp]
    \centering
    \begin{subfigure}{0.9\columnwidth}
        \centering
        \includegraphics[width=0.9\textwidth]{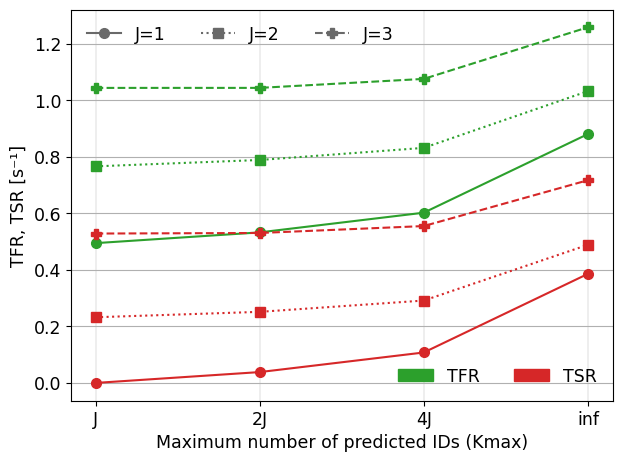}
        \caption{TFR and TSR $\downarrow$ [$s^{-1}$]}
        \label{fig:exp2_tsr}
    \end{subfigure}
    \begin{subfigure}{0.9\columnwidth}
        \centering
        \includegraphics[width=0.9\textwidth]{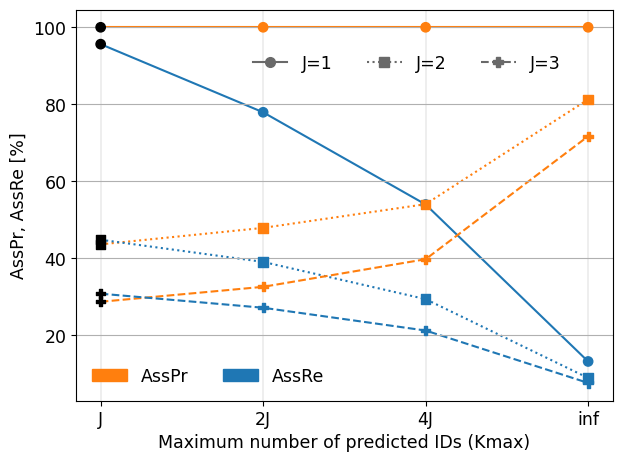}
        \caption{AssPr and AssRe $\uparrow$ [\%]}
        \label{fig:exp2_assprasspre}
    \end{subfigure}
    \caption{Mean identity tracking metrics using PF tracker on the three subsets of \hbox{LibriJump} dataset, for different $K_{max}$ values. Black markers in Fig.~\ref{fig:exp2_assprasspre} is for highlighting $K_{max} = J$ scores.}
    \label{fig:exp2}
\end{figure*}

Single speaker metric scores using PF tracker can be found on Fig.~\ref{fig:exp1} (\hbox{LOCATA} dataset), solid lines of Figs.~\ref{fig:exp2_tsr} and~\ref{fig:exp2_assprasspre} (\hbox{LibriJump}-1spk). The \textbf{1spk} line on Tab.~\ref{tab:tab_exp2} gives NN tracker's performances. 
AssA follows similar evolution to AssRe and is not displayed.

On both Figs.~\ref{fig:exp1_tfr_tsr} and~\ref{fig:exp2_tsr}, TFR is higher than TSR, since it counts the number of broken and swapped tracks, in comparison with TSR which measures only the swaps. 
For NN tracker, they are both equal to 0.43 $s^{-1}$ in the 1spk subset. This indicates the absence of broken tracks, linked to excellent detection performance (broken tracks are linked to the presence of FNs).

AssPr is near 100\% for all datasets and trackers. This is expected in the single speaker case with only one ground truth : merging errors are impossible, which leads to high AssPr.
Thus, the only possible identity assignment error is the creation of redundant IDs for the same ground truth, as measured by \hbox{AssRe}.

On \hbox{LOCATA} dataset, we compare identity assignment performances given continuous (Tasks 1 and 3) and discontinuous (Track 3*) tracks. We focus on AssRe and TSR, as AssPr is optimal and TFR is influenced by detection errors and not only association errors.
On continuous tracks, PF tracker displays optimal performance in term of TSR for all recordings of Task 1, and 
4 out of 5
recordings of Task 3, showing the ability of PF tracker to maintain a consistent predicted ID when the single speaker track is continuous. 
Given discontinuous tracks (Task 3*), performance degrades so that only 
1 out of 5
recordings remains optimal. 
Similar conclusions can be drawn from AssRe, since it measures the tendency to attribute new IDs, like \textit{swaps} in TSR. Although, it does provide complementary information due to its computation at the global level.
For example, in Task~3*, the worst recording has a TSR of 0.28~$s^{-1}$ and AssRe of 47\%. The TSR score translates into an ID change occurring approx. every 3~seconds, while the AssRe means that half the ground truth track is split across other predicted tracks. 
This gives a complementary understanding of PF tracker's behavior.

\subsubsection{Multiple speaker scenes analysis}

We now focus on more complex multi-speaker scenes, namely \hbox{LibriJump}-2spk and 3spk. PF tracker's results on these subsets are plotted on Fig.~\ref{fig:exp2} (dotted lines), and those of NN tracker are on the corresponding lines of Tab.~\ref{tab:tab_exp2}. 

We first analyze the behavior common to both PF and NN tracker.
Considering AssPr and AssRe, when \hbox{$K_{max} = J$} (\textbf{2spk} line for NN tracker and black markers on Fig~\ref{fig:exp2_assprasspre} for PF tracker), we notice that they are approximately equal for both trackers. 
This highlights the trackers limitations in term of ID assignment in such cases, and their tendency to interchange between same IDs over time.
As a result, any counted FPA for a given TP is likely to be counted as a FNA for another TP, which balances AssPr and AssRe.
Such behavior is not visible when looking at TSR and TFR, since TSR looks at when an ID change occur without further specifications, and TFR informs about the presence of FNs.

Looking now at NN tracker, TFR and TSR are equal on the 2spk subset, for the same reasons as in the single speaker case.
On the 3spk subset, TFR becomes superior to TSR, which underlines the presence of FNs since NN tracker has only has 2 outputs for localizing 3 speakers.
Looking at association metrics, AssPr (36.6\%) is largely inferior to its corresponding AssRe (47.3\%). This occurs for the first and only time and corresponds to the only case where $K_{max} < J$ in our experiments. 
With a number of possible ID predictions inferior to the number of ground truths, NN tracker is forced to merge multiple ground truths into one prediction, which lowers AssPr.

For PF tracker, similar to the single speaker case, TFR is superior to TSR on both 2spk and 3spk subsets. Also, they worsen when $K_{max}$ augments, showing the increase of swaps as the tracker has more freedom in ID assignment.
Interestingly, Fig.~\ref{fig:exp2} shows that AssRe and TSR degrade and AssPr improves when $K_{max}$ augments.
As PF tracker has more flexibility in ID assignment, it becomes more likely to initiate new tracks upon discontinuities, rather than maintaining the same ID for longer periods of time. This results in splitting a track into a multitude of predictions, which lowers AssRe, as well as TSR since each new ID change causes a swap. AssPr, linked to the opposite tendency to merge multiple tracks into a single predicted one, increases.

\begingroup
\setlength{\tabcolsep}{5.2pt} 
\renewcommand{\arraystretch}{1} 
\begin{table}[!htbp]
\centering
\caption{Tracking metrics on the three subsets of \hbox{LirbiJump} dataset using NN tracker ($K_{max} = 2$).}
\label{tab:tab_exp2}
\begin{tabular}{cccccccc}
\hline
   \multirow{2}{*}{Subset}   & & TSR & TFR &  & AssA & AssPr & AssRe  \\ 
\cline{3-4} 
\cline{6-8} 
  & & \multicolumn{2}{c}{[$s^{-1}$] $\downarrow$} & & \multicolumn{3}{c}{[\%] $\uparrow$} \\ \hline
   \textbf{1spk} & & 0.43 & 0.43 &  & 55.4 & 100 & 55.4 \\
   \textbf{2spk}  & &  0.87 & 0.87 & & 38.1 & 54.0 & 54.2 \\
   \textbf{3spk}  & & 0.84 & 1.15 & & 26.6 & 36.6 & 47.3 \\ \hline
\end{tabular}
\end{table}
\endgroup

\section{Conclusion}

In this paper, we presented the problem of tracking intermittent and moving speakers, and the challenges raised movement during silence, in term of spatial discontinuity.
We presented \hbox{LibriJump}, a synthetic dataset of acoustic scenes containing speakers with discontinuous spatial tracks, to promote further interest to this seldom explored, yet realistic scenario.
We gave an overview of current tracking metrics used in sound source tracking, and proposed to complement them with tracking association metrics from computer vision. 
We finally analyzed the identity assignment performance of two tracking systems upon continuous and discontinuous tracks, and showed the complementarity of proposed and existing metrics.

\printbibliography[heading=bibnumbered, title={References}]

\end{document}